\documentclass[journal]{IEEEtran}
\usepackage{graphicx} 
\usepackage{color}
\usepackage[rightcaption]{sidecap}
\usepackage[]{algorithm2e}
\usepackage{wrapfig}
\usepackage[noadjust]{cite}
\usepackage{filecontents}
\usepackage[utf8]{inputenc}
\usepackage{dirtytalk}

\ifCLASSINFOpdf

\else

\fi
\begin{document}

\title{Perspective Paper: TickTalk - Timing API for Dynamically Federated Cyber-Physical Systems}

        \author{\IEEEauthorblockN{Bob Iannucci\IEEEauthorrefmark{2}, Aviral Shrivastava\IEEEauthorrefmark{1}
 and Mohammad Khayatian\IEEEauthorrefmark{1}}
\IEEEauthorblockA{\\
\IEEEauthorrefmark{2}Carnegie Mellon University,
\IEEEauthorrefmark{1}Arizona State University\\
bob@sv.cmu.edu,
aviral.shrivastava@asu.edu,
mkhayati@asu.edu
\vspace{-0.5cm}
}
}

\maketitle

\begin{abstract}
Although timing and synchronization among a dynamically-changing set of sensing, computing, and actuating elements and their related power considerations are essential to many cyber-physical systems (CPS), these concepts are absent from today’s programming languages, forcing programmers to handle these matters outside of the language and on a case-by-case basis. This paper proposes a framework for adding time-related concepts to languages. Complementing prior work in this area, this paper develops the notion of dynamically federated islands of variable-precision synchronization and coordinated entities through synergistic activities at the language, system, network, and device levels. At the language level, we explore constructs that capture key timing and synchronization concepts.  At the system level, we propose a flexible intermediate language that represents both program logic and timing constraints together with run-time mechanisms. At the network level, we argue for architectural extensions that permit the network to act as combined computing, communication, storage, and synchronization platform.  At the device level, we explore architectural concepts that can lead to greater interoperability, the easy establishment of timing constraints, and more power-efficient designs.
\end{abstract}


\section{Introduction}
\label{sec:intro}
Our imagination and concepts of Cyber-Physical Systems (CPS) are transforming our vision of the Internet of Things (IoT), Internet of Everything (IoE) and so-called smart cities. The concepts are simultaneously appealing and puzzling. The appeal comes from the ability to apply computing and communications technologies in numerous ways and on a wide scale to improve life. The puzzling aspect is how to achieve it. If we can sense anything, and actuate anything, what useful things can we do? One example is to empower anyone to track people and things valuable to them (their child on a bicycle, a truck, stolen property) using information gleaned from a smart city’s collective pool of sensors and to initiate some appropriate actions. There are many similar time-sensitive, distributed computing tasks in a smart city or other IoT networks that involve interaction with spatially distributed nodes \cite{weiss2015time}. Many such applications written by various programmers should be able to share the city-wide CPS infrastructure. Thus each CPS node will accept snippets of code from separately-created programs to run in a coordinated way with other nodes in the system. For instance, one programmer may be interested in taking a picture at 4:00 pm while another programmer is interested in sensing the temperature in the environs of the same CPS node at 4:00 pm. How do we make all this possible, especially when the programs are being developed separately and without coordination? How do we know, for example, if the combined functionality is even possible? How do we make programming these geographically distributed time-sensitive systems easier? Programming CPS is difficult because it combines the complexities of distributed programming and time-sensitive programming -- both of which bring portability and scalability issues \cite{shrivastava2016time}. What is a good distributed-timing application programming interface (API) that can ease this burden? Can such an API offer clean semantics for reasoning about the time-related behavior of these programs? In this paper, we explore these questions.

\section{Need for Timing and Synchronization API}
\label{sec:Need}
\begin{figure}[htb]
\centering
\includegraphics[width=\columnwidth]{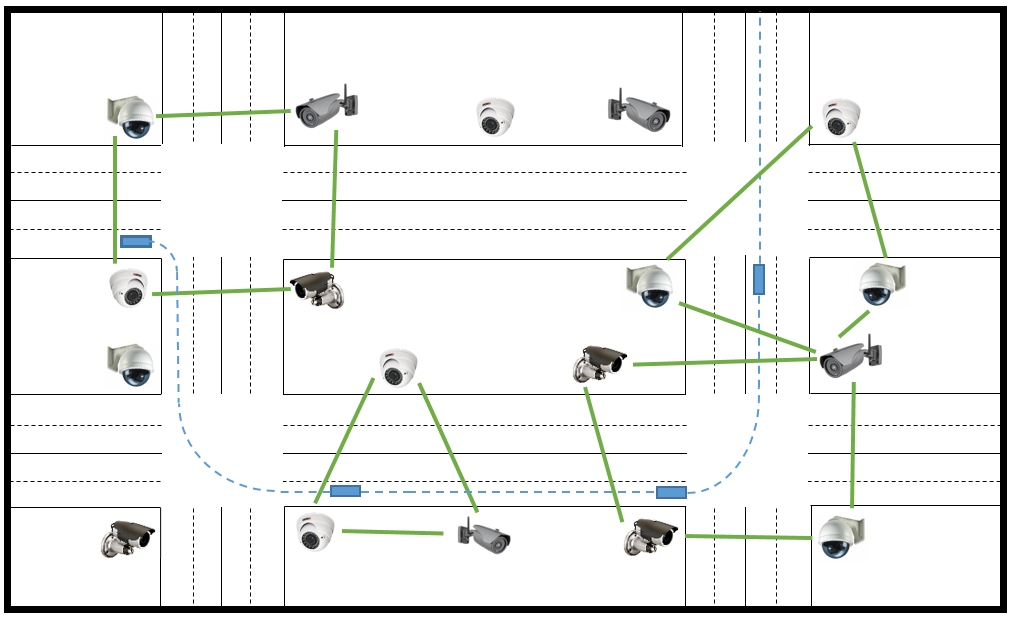}
\caption{Dynamically enrolling separately-managed cameras scattered around a city to track a truck}
\label{fig:DFCPS}
\end{figure}

Consider the example of a smart city in which a transportation company wants to ``observe'' one of its assets, in this case, an en-route truck. Imagine that they have the authority to dynamically recruit pools of separately-installed and -managed cameras that may be found around the smart city (\emph{e.g.,} on buildings, poles) to get a 3D video view of the truck’s movement. These scattered CPS nodes can be fused to become a federated cyber-physical system (FCPS) which can sense, compute, communicate and actuate as an integrated whole. But even the simple notion of collecting video from cameras near the truck will involve enrolling more cameras over time as the truck moves. We call such a system a \emph(dynamically) federated cyber-physical system (DFCPS). Figure~\ref{fig:DFCPS} depicts the truck moving around the city and how the notion of ``nearby cameras'' must evolve. Network boundaries between sets of separately-managed cameras are depicted with solid green lines.  The dotted blue line shows the trajectory of the truck.

\begin{figure*}[t]
\newenvironment{boxed}
    {\begin{center}
    \begin{tabular}{|p{0.80\textwidth}|}
    \hline\\}
    {\\\\\hline
    \end{tabular} 
    \end{center}}
    \begin{boxed}
    \textbf{loop\{}\hfill \\
    \qquad {\fontfamily{qcr}\selectfont\small// assume (x,y) is the predicted position of the object} \hfill \\
    \qquad $S$ = getSensors($x,\ y$, 100 \textit{m}); {\fontfamily{qcr}\selectfont\small// get sensors within 100 meters of (x,y)}\hfill \\
    \qquad $A$ = emptySet(sizeOf($S$)); {\fontfamily{qcr}\selectfont\small// empty set of images}\hfill \\
    \qquad withSynchronization($S$, 1 \textit{us}, self) \{\hfill \\
    \qquad \qquad{\fontfamily{qcr}\selectfont\small // within this block, the sensors will synchronize to 1 us accuracy}\hfill \\
    
    \qquad \qquad $a$ = simultaneously($S$.captureImage());\hfill \\
    \qquad \}\hfill \\
    \qquad 3DImage = create3DImage($A$);\hfill \\
    \qquad 3Dmodel.addImage(3DImage);\hfill \\
    \qquad ($x',\ y'$) = predictNextPosition($x,\ y,\ A,\ t$); {\fontfamily{qcr}\selectfont\small// new predicted position} \hfill \\
    \qquad if (($x',\ y'$) == ($x,\ y$)) break; else ($x',\ y'$) = ($x,\ y$);\hfill \\
    \} {\fontfamily{qcr}\selectfont\small // end loop} \hfill 
    \end{boxed}
    \caption{Pseudo-code for tracking a moving object using scattered cameras in the city.}
    \label{fig:code1}
    \vspace{-0.5cm}
\end{figure*}

One approach to dynamic federation is to impose some sort of hierarchy -- within co-located clusters, one node can serve as a leader to be responsible for in-cluster synchronization and global communication. This leader could arrange for the other nodes to take pictures at specified times. The leader gathers these frames and constructs a dynamic 3D model from which estimates of the future position of the object can be made. 

The typical approach would be to write application-specific code for each participating node together with code to coordinate the actions of these nodes. Verification and validation of separate applications working together are challenging because it necessitates the specification of the overall system's behavior. Testing is likewise challenging. A better approach and, we dare say, one that might be more acceptable to the millions of programmers who might invest their efforts in the creation of such smart city apps, is to develop one \emph{integrated} application that can be verified and then, separately, decomposed and distributed to the nodes on which it will run. A representative pseudocode is shown in Figure~\ref{fig:code1}. While the pseudocode may seem simple, it highlights important opportunities and challenges that emerge from the very nature of programming a geographically-distributed aggregate of computing resources. In this section, we will discuss challenges of adding timing concepts to the programming language as well as achieving synchronization for a scattered time-sensitive system.

\subsection{Time-related Programming and Synchronization}
To achieve deterministic timing on IoT/CPS devices, timing must be made a correctness criterion and not just a performance factor. Hence, by making timing constraints \cite{mehrabian2017timestamp} and requirements part of the formal model/program, reasoning about and verifying timing requirements becomes possible. Correctly defined, the semantics of timing primitives in specification models determine whether correctness properties can be checked by inherent construction, symbolic analysis, explicit simulation, or only in the implementation. However, as long as the timing specification is not a part of the programming language, whether a system implementation meets the timing requirements or not, can only be checked by testing after building the whole system. 

Programmers of the future will have to make the system achieve correct timing even though today’s popular languages lack mechanisms for expressing the needed time-related concepts. For example, in C, we lack the following concepts:
\\ \\ {\fontfamily{qcr}\selectfont
printf(''hello world $\backslash$n'', @4:35 PM);}\\ \\
Even if programmers had such expressive power, making good on their intent will require new mechanisms in the underlying network and devices. For applications similar to our example, nearby CPS nodes only need to be synchronized among themselves and do not need to be synchronized to the time server nor to coordinated universal time (UTC). We simply need to create the conditions under which they all take photos essentially at the same instant. 

Other applications may need synchronization to an external reference (\emph{e.g.,} UTC). For instance, a user may be interested in polling data at exactly 11:00 AM. Therefore, all sensing/actuating nodes of FCPS must have a common understanding of the real time and must execute the sensing/actuating code exactly at the specified time regardless of worst-case execution time (WCET) of the underlying computation platform \cite{Wilhelm:2008}, network delay, local clock drift, and so on. Since time-related concepts are absent from today programming languages, programmers must handle these matters outside of the language.

\subsection{Cost-Power Efficiency}
As sensors proliferate in a smart city, the cost of providing each one with a wired power connection will become overwhelming \cite{Iannucci2017-CSS}. Devices that operate for years on batteries and/or harvest energy will be preferred. A very closely related issue for small in-the-environment sensors and actuators is the power-cost of achieving time awareness. Time synchronization between two or more devices necessitates frequent communication. As the need for precision increases, so does the power-cost of achieving it. Further, we must accept that energy-constrained devices must be \emph{mostly off}, implying that such devices either need to invest power in precise, low-drift internal clocks or plan to wake up ``just in time'' to perform over-the-network resynchronization.  It is likely to be power-advantageous for nodes to continuously be loosely synchronized to UTC--with just enough power put into the local clock to enable in-time wakeup, wireless synchronization, and the on-time capturing of the picture (after which time-keeping can revert to low precision).



It will be important to develop power-efficient solutions for implementing the behind-the-scenes (runtime) mechanisms that will reliably achieve application-level timing precision at the lowest possible power levels and without introducing jitter or timing errors.   


\subsection{Support for Heterogeneity}
Returning to our truck-tracking example, it unlikely that the \textit{borrowed} cameras will be identical--same vendor, same programming interface, same functionality, same performance. Rather, in the information-sharing economy of this smart city, cameras and other sensors are likely to be dissimilar.  It is already difficult enough to imagine fusing the data, but DFCPS compounds the programmer's challenges by requiring the management of timing across an array of dissimilar devices. This is an uncommon, rather than a common, software engineering skillset. We are, then, left with the conundrum that the domain (DFCPS) requires programmer-management of time, yet programmers and their tools (languages, compilers, run-time systems) are ill-prepared for this. Viewing this as an architectural problem, we seek a solution that only requires that the programmer specify the timing \emph{intent} (\emph{do these three things at the same time}) and leaves to the run-time mechanisms the realization of these requirements.  As we will see, this necessitates some minimal augmentation of the timing mechanisms in computing and communications equipment. Precision protocols for network-based transport of time information (\emph{e.g.,} PTP/IEEE-1588) similarly identify the need for specific hardware support.



\subsection{Multi-Tenancy: Code Block Multiplexing}
Our tracking example and others suggest that significant value will be derived from recruiting sensors/actuators dynamically, and making them sense or take action in a synchronized fashion. As the value of the smart city catches on, our programmer won’t be the only one using the cameras. Many apps in this smart city will likely want to concurrently share some or all of the cameras. At that point, our carefully-synchronized application will be faced with the challenge of sharing the hardware with other concurrent applications. While sharing an IoT device (\emph{e.g.,} a motion sensor, a camera) across applications may seem simple, different applications will impose differing and possibly conflicting requirements with respect to time. At a minimum, we imagine the need to discover potential conflicts, harmonize them when possible and signal irreconcilable conflicts otherwise.  

As a starting point, if programmer-derived timing requirements are expressed cleanly and clearly, we can imagine compiling and run-time tools that will enable this sorting-out of separately-created timing constraints that come together on a single hardware device. But there are subtle complications.  Imagine that code block $b1$ seeks to run on a given node (say, one of our cameras) and, per the programmer's intent, it is synchronized to some reference clock $c1$.  But along, comes code block $b2$ from a different application that also seeks to run on this same camera. Alarmingly, its programmer has elected to use a different and possibly incompatible (with $c1$) reference clock $c2$. These clocks may differ in frequency, phase and/or epoch--and for good reasons known to their programmers. In this case, we must go beyond considerations of simply sharing the computing resources (by traditional virtualization techniques, for instance) to embracing the notion that clocks themselves must be virtualized--allowing for a separate clock per code block. The run-time system must somehow deal with issues of non-synchronization among these reference clocks.


\begin{figure*}[htb]
\centering
\includegraphics[width=1.8\columnwidth]{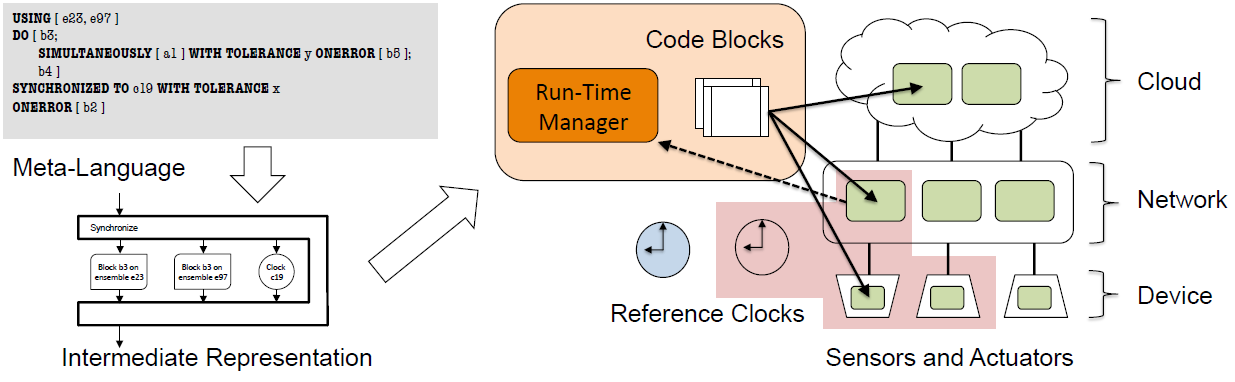}
\caption{The proposed architecture. High-level programs are decomposed to an intermediate-level form in which time-based operations are explicitly represented. }
\label{fig:overall}
\vspace{-0.5cm}
\end{figure*}

\section{Our Approach}
\label{sec:our approach}
Our approach advances the concept of an easily-programmed federated cyber-physical system (FCPS) that hides the inherent complexities of synchronization of distributed actions. We model a FCPS as a tuple \textit{(C, E, B)} in which:
\\
$C= \{c_1,c_2, ...\}$ is the set of reference clocks. each clock is characterized by its frequency, phase, jitter, etc.
\\
$E=\{e_1,e_2,...\}$ is the set of computing, storage, actuating and sensing ensembles (think: nodes.  More on this below.), each of which has its own set of local clocks.
\\
$B=\{b_1,b_2,...\}$ is the set of computational blocks (program fragments) within which, actions can be scheduled to take place at specific times.

We use the term \emph{ensemble} to capture the notion of an element that has computing, storage, communication and timing capabilities that allow it to accept one or more code blocks. It is worth noting that our notion of ensemble is intentionally broad and is intended to abstract the hardware for sensor and actuator nodes (including the computing, storage, and communication chips associated with them), network-resident computing facilities such as would be necessary to implement fog computing or cloudlets, and cloud computing equipment such as would be found in large, virtualized data centers. We specifically contemplate the additional possibility of dynamically migrating code blocks from ensemble to ensemble, implying a notion of common base functionality. We use the term ensemble instance to denote an invocation of a code block on a particular ensemble with a particular ensemble-local clock. 

By characterizing ensembles in this way, we enable the possibility of taking a single program and breaking it into pieces that run concurrently in the cloud, in the network, and the devices. Note that this is different than the traditional model in which the cloud code is written by one team, the device code is written as part of the development of a power-constrained embedded system, and the network is largely un-programmable by non-specialist developers. We imagine, as a possible outcome of this research, the creation of a reference architecture for ensembles that could aid in assuring growing interoperability among future smart city elements. 

As depicted in Figure~\ref{fig:overall}, programs written in a suitable high-level language with constructs will be translated into a dataflow graph. Dataflow provides a clear dependency-driven graph interpretation framework to which we seek to add reference clock synchronization semantics. A simple formulation is to decompose FCPS meta-programs into graphs in which each node represents the instance of a code block on a specific ensemble. Synchronization and simultaneity dependencies to reference clocks can be explicitly represented. Synchronization, when established, will yield tokens that become part of the firing rules for the respective nodes, while simultaneity has to be ensured by programming and analysis. Ensembles are depicted in green. As an example, a sub-domain involving a network ensemble, a sensor ensemble, and an actuator ensemble is highlighted in red. System operations such as code block placement and synchronization are handled by the Run-Time Manager (RTM). Feedback from network elements to the RTM facilitate improved synchronization (dashed arrow).

At the language level, timing semantics for functionality that are commonly used can be categorized as \cite{mehrabian2017timestamp}:

\noindent\textbf{Frequency-based sensing/actuating.} Utilization of periodic actions is very common in IoT applications and is characterized as ``take an action every $x$ nano, Micro or Milliseconds.'' Certain frequencies of sensing or actuating are required to achieve desired Quality of Control (QoC).

\noindent\textbf{Syntonization and Synchronization.} Certain levels of synchronization are required for many applications, however, high precision time synchronization in a large scale system will cause network traffic and consequently network delay is less predictable. This problem can be addressed by defining variable synchronization levels for ensembles.

\noindent\textbf{Simultaneous sensing/actuating.} Performing two or more concurrent actions is very common in Multi-Agent Systems which are widely used for different purposes like Distributed Learning and Problem Solving, Decentralized Control, Formation Control, and the like. Hence, as a functionality, the application must push code blocks into ensembles so that desired actions be taken simultaneously. 

\noindent\textbf{Latency-based sensing/actuating.}
In time-sensitive applications, sensor information and results computed from the sensors are valid for only a specific temporal interval, necessitating bounded or fixed latency constraints on communication and computation. Timeliness, or the temporal limits of the application to communicate information or execute an action can be described through \textit{latency-based} specifications. 

At the network level, in assimilating information across a smart city, the nanosecond-scale of computation is dwarfed by the tens- to hundreds of milliseconds needed to traverse network connections across a city. The worst-case round-trip time for a cyber-physical control loop (sensing, computing, and acting) can easily exceed 1000 milliseconds, making classical cloud-based cyber-physical systems useless for cases requiring response times in the deep sub-second regime. One important and promising approach to reduce CPS latency when mobile networks are involved is moving the computation into the network itself to expose the trade-off between the network latency, amount of computation on end-devices, and the network bandwidth requirements. \textit{Cloudlets} \cite{satyanarayanan2014cloudlets} and \textit{fog computing} \cite{bonomi2012fog} have motivated research in this area. 

While programmer specification of timing requirements is necessary, it is not sufficient. The nodes and network impose constraints. As such, we seek to extract information about both latency and latency variability in real time from the FCPS and to feed this information in a usable form back to the programmer. We argue that most realistic networks exhibit time-varying behavior and that knowledge of the current state of the network can be used in dynamically optimizing how a distributed program works.

\section{Acknowledgement}
This material is based upon work supported by the National Science Foundation under grants no. CPS 1646235 and CPS 1645578.

\bibliographystyle{IEEEtran}
\bibliography{main.bib}

\end{document}